%% file: ms.tex
\documentclass[preprint2]{aastex}
\shorttitle{Radio Variability of VeLLOs}
\shortauthors{Choi et al.}

\usepackage{times}
\slugcomment{To appear in the Astrophysical Journal}

\begin{document}

\fontsize{10}{10.6}\selectfont

\title{Radio Variability Survey of Very Low Luminosity Protostars}
\author{\sc Minho Choi$^{1}$, Jeong-Eun Lee$^2$, and Miju Kang$^{1,3}$}
\affil{$^1$ Korea Astronomy and Space Science Institute,
            776 Daedeokdaero, Daejeon 305-348, Korea;
            minho@kasi.re.kr \\
       $^2$ Department of Astronomy and Space Science,
            Kyung Hee University, Yongin, Gyeonggi 446-701, Korea \\
       $^3$ Max-Planck-Institut f{\"u}r Radioastronomie,
            Auf dem H{\"u}gel 69, D-53121 Bonn, Germany}

\begin{abstract}
Ten very low luminosity objects
were observed multiple times in the 8.5 GHz continuum
in search of protostellar magnetic activities.
A radio outburst of IRAM 04191+1522 IRS was detected,
and the variability timescale was about 20 days or shorter.
The results of this survey and archival observations suggest
that IRAM 04191+1522 IRS is in active states about half the time.
Archival data show that L1014 IRS and L1148 IRS were detectable previously
and suggest that at least 20\%--30\% of very low luminosity protostars
are radio variables.
Considering the variability timescale and flux level of IRAM 04191+1522 IRS
and the previous detection of the circular polarization of L1014 IRS,
the radio outbursts of these protostars
are probably caused by magnetic flares.
However, IRAM 04191+1522 IRS is too young and small
to develop an internal convective dynamo.
If the detected radio emission is indeed coming from magnetic flares,
the discovery implies
that the flares may be caused by the fossil magnetic fields
of interstellar origin.
\end{abstract}


\section{INTRODUCTION}

Magnetosphere is an integral part of a star
and plays important roles in the dynamics of ionized gas
in and around the star (Donati \& Landstreet 2009; Mestel 2012).
The magnetosphere manifests itself
through powerful outbursts of energy such as radio and X-ray flares,
and such magnetic activities start
well before the onset of stable hydrogen burning
(Feigelson \& Montmerle 1999).
(In this paper, we use the term ``outburst''
to mean a sudden increase of brightness
and the term ``flare''
to mean an outburst event caused by magnetic reconnection.
This flare is not necessarily a solar-type flare.)
Many protostars drive well-collimated outflows (Arce et al. 2007),
and the outflows are generated
through magnetocentrifugal ejection mechanism (Pudritz et al. 2007),
which suggests that protostars may possess well-organized magnetic fields.
The magnetic fields of young stellar objects (YSOs)
may be seeded by the interstellar magnetic fields in the natal cloud
(Mestel 1965; Tayler 1987).
However, it is difficult to study the origin of stellar magnetic fields
and to obtain observational evidence
for the magnetic fields of young protostars,
because Class 0 protostars,
objects in the earliest stage of stellar evolution (Andr{\'e} et al. 1993),
are surrounded by opaque layers of gas
that block the radio and X-ray emission from the view of outside observers
(Feigelson \& Montmerle 1999).
Especially, the partially ionized winds of typical protostars
are opaque to radio waves (Andr{\'e} 1996).

The free-free optical depth of a protostellar wind can be roughly estimated
from the mass accretion rate.
For a typical protostar, the mass accretion rate is 
$\sim$2 $\times$ 10$^{-6}$ $M_\odot$ yr$^{-1}$ (Shu et al. 1987).
Assuming an accretion-to-outflow conversion factor of 10\%
and an ionization fraction of 1\%,
the mass-loss rate of ionized wind is
$\sim$2 $\times$ 10$^{-9}$ $M_\odot$ yr$^{-1}$.
Using Equation (5) of Andr{\'e} et al. (1988),
and assuming a wind speed of 200 km s$^{-1}$,
an emission-region radius of 3--9 $R_\odot$
(1--3 times the protostellar radius),
and a wind temperature of 10$^4$ K,
the optical depth would be 400--10,000 at 8.5 GHz.
Therefore, magnetic flares of typical protostars
are unobservable in the radio wavelength band.

Recently discovered low-luminosity protostars
present a new opportunity
to directly observe the protostellar atmospheres.
These very low luminosity objects (VeLLOs)
were revealed by observations in infrared wavelengths
and have internal luminosities less than 0.1 $L_\odot$
(Young et al. 2004; Dunham et al. 2006).
VeLLOs provide interesting insights
into the physics and chemistry of low-mass star formation
(Lee 2007; Dunham et al. 2008).
They may have smaller mass accretion rates, hence smaller mass-loss rates,
than typical protostars.

The internal luminosities of VeLLOs are in the range of 0.02--0.1 $L_\odot$
(Dunham et al. 2008).
For a protostar of $\sim$0.075 $M_\odot$ (the stellar--substellar boundary)
with a radius of 3 $R_\odot$,
the mass accretion rate would be
0.3--1.3 $\times$ 10$^{-7}$ $M_\odot$ yr$^{-1}$.
Using the assumptions above,
the free--free optical depth of a VeLLO wind would be 0.08--40.
Therefore, the magnetic flares of VeLLOs are detectable,
especially when observed during powerful and large-size flare events.

Much of the knowledge about the magnetic fields of YSOs
comes from the radio and X-ray observations of more evolved objects
such as T Tauri stars, Class I protostars,
and the late-stage protostar R CrA IRS 5
(Andr{\'e} 1996; G{\"u}del 2002; Forbrich et al. 2006;
Choi et al. 2008, 2009; Hamaguchi et al. 2008).
These objects showed that one of the key characteristics
is the flux variability of emission from nonthermal electrons.
The variability timescale of magnetic flares of YSOs is usually hours to days.

In this paper, we present the results of our monitoring survey of VeLLOs
using the Very Large Array (VLA)
of the National Radio Astronomy Observatory (NRAO).
We describe our observations and archival data in Section 2.
In Section 3, we report the results.
In Section 4, we discuss the magnetic activities of protostars.

\section{OBSERVATIONS AND DATA}

\subsection{Source Selection}

We observed the VeLLOs in the list compiled by Dunham et al. (2008).
Out of the 15 VeLLOs in the list,
10 objects of high declination were selected.
These sources stay over the observable elevation limit
for relatively long durations
and are suitable for a survey with VLA.
The selected sources are listed in Table 1.
The target sources are protostellar objects
in nearby star-forming molecular clouds.
The sources are divided into two groups depending on the right ascension
so that all the sources in a group
can be observed in a single observing run.

\input{tab1.tex}

\subsection{VLA Observations}

The survey targets were observed using VLA in 2009
in the standard $X$-band continuum mode (8.5 GHz or $\lambda$ = 3.5 cm).
The total bandwidth for the continuum observations was 172 MHz
(43 MHz for each of the four intermediate-frequency channels).
The observations were carried out in the D-array configuration
toward the infrared source positions determined
using the {\it Spitzer Space Telescope} (Dunham et al. 2008).
Each target source was observed three times with an interval of $\sim$20 days
(Group A sources on October 18, November 8, and November 30,
and Group B sources on October 31, November 8, and November 30)
so that the flux variability can be determined
if the source went through a radio outburst.
In each observing run,
the on-source integration time for each target source was $\sim$12 minutes.

The phase was determined by observing nearby quasars.
The flux was calibrated by setting the flux densities
of the quasars 0542+498 (3C 147) and 1331+305 (3C 286) to 4.7 and 5.2 Jy,
respectively.
Maps were made using a CLEAN algorithm.
With a natural weighting,
the 8.5 GHz continuum data produced synthesized beams of $\sim$9$''$ in FWHM.

\subsection{VLA Archival Data}

In addition to the survey data,
several VLA data sets retrieved from the NRAO Data Archive System
were also analyzed.
We searched for 8.5 GHz continuum observations of the target regions
and found 10 data sets.
All of them have a continuum bandwidth of 172 MHz.
These data sets were analyzed
following the standard VLA data reduction procedure.
The archival data by themselves are not adequate
for the confirmation of magnetic flares
because their time resolution is worse than a year.

\section{RESULTS}

In our survey, only DCE 1 was detected.
For each of the undetected sources,
data from all the three observing runs were combined
to reduce the noise,
but no additional detection was made.
In the archival data sets, DCE 1, 32, and 38 were detected,
indicating that they are variable radio sources.
The results are summarized in Tables 2 and 3.

\input{tab2.tex}
\input{tab3.tex}

\subsection{IRAM 04191+1522 IRS (DCE 1)}

DCE 1 corresponds to IRAM 04191+1522 IRS,
a Class 0 protostar driving a highly-collimated bipolar outflow
(Andr{\'e} et al. 1999; Dunham et al. 2006).
IRAM 04191+1522 IRS was detected clearly
in the first observing run (Figure 1).
The position of the 8.5 GHz continuum source was
$\alpha_{2000}$ = 04$^{\rm h}$21$^{\rm m}$56$\fs$79
and $\delta_{2000}$ = 15$^\circ$29$'$45$\farcs$5,
which agrees with that of the infrared source within 1$\farcs$4.
Considering the 9$''$ beam size,
the radio source is positionally coincident with the infrared source.
Circular polarization was not detected.
From the noise level in the Stokes-$V$ map,
the 3$\sigma$ upper limit on the polarization fraction is 21\%.

\input{fig1.tex}

IRAM 04191+1522 IRS weakened in the subsequent observing runs (Figure 1).
Figure 2(a) shows the light curve.
While IRAM 04191+1522 IRS is clearly a variable radio source,
the timescale of the variability in 2009 October--November
can be constrained only roughly
because the number of data points is small.
In the simplest case,
the light curve may be showing a single outburst event.
If the peak intensity occurred before the first run,
the light curve in Figure 2(a) covers the waning phase only,
and the decay half-life is $\sim$20 days.
If it occurred between the first and second runs,
the FWHM timescale may be in the range of 10 to 40 days.
(The 40 day limit is for a box-shaped light curve,
and a more likely value for Gaussian-like profiles is $\sim$20 days.)
More realistically, the light curves of YSO radio outbursts show
month-scale periods of enhanced activities
that consist of several day-scale outbursts
(Bower et al. 2003; Forbrich et al. 2006).
If this is the case for IRAM 04191+1522 IRS,
and if the detections of the first and second runs were
from separate outburst events,
the timescale of each event may be shorter than $\sim$10 days.
In summary, the timescale of the IRAM 04191+1522 IRS flux variability
may be $\sim$20 days or shorter.

\input{fig2.tex}

The data over 17 yr from 1992 to 2009 show
that IRAM 04191+1522 IRS was detectable ($>$ 0.05 mJy)
about 50\% of the time (Table 2, Figure 2(b)),
which suggests
that the radio outburst is not a rare phenomenon on this protostar.
The 1996/1997 observations and results
were reported previously by Andr{\'e} et al. (1999).
In all the four runs when IRAM 04191+1522 IRS was detected,
the source was unresolved.

The radio emission of IRAM 04191+1522 IRS in the quiescent state
can be constrained from the observations of 2003 January.
The 3$\sigma$ upper limit of the quiescent flux density is 0.02 mJy.
Then the 2009 October--November outburst
was stronger than the quiescent level by a factor of seven or larger.
If we limit our attention only to the 2009 observations covering 43 days,
the peak flux (first run) is $\sim$3.4 times the upper limit of the third run.

\subsection{L1148 IRS (DCE 32)}

DCE 32 corresponds to L1148 IRS,
a Class I protostar driving a compact outflow
and residing in a low-mass cloud core (Kauffmann et al. 2011).
L1148 IRS was detected in 2005 (Table 3).
The radio source position agrees with that of the infrared source
within 0$\farcs$9.
L1148 IRS was undetected in our survey.
L1148 IRS seems to be a time-variable source (Figure 3),
but it is difficult to understand the nature of this radio source
because it was weak and detected only once.

\input{fig3.tex}

\subsection{L1014 IRS (DCE 38)}

DCE 38 corresponds to L1014 IRS,
a Class 0 protostar driving a compact outflow
(Young et al. 2004; Bourke et al. 2005).
L1014 IRS was detected in 2004 and 2005 (Table 3).
The observations and results were reported by Shirley et al. (2007)
along with the information on the detections at 4.9 GHz.
The radio source position
agrees with that of the infrared source within 0$\farcs$7.

As L1014 IRS was not detected in 2009,
it is clearly a time-variable source (Figure 3).
The detections in 2004 and 2005 may represent separate radio outbursts,
and the outburst timescale of each event cannot be constrained.
Alternatively, it is in principle possible
that the 2004/2005 detections were from a single event
with a timescale longer than 1.4 yr.
In 2004, circular polarization of the 8.5 GHz continuum
was marginally detectable
with a Stokes-$V$ flux density of 0.030 $\pm$ 0.008 mJy.
The polarization fraction is $\sim$30\%.
Shirley et al. (2007) reported
that the 4.9 GHz continuum flux is also variable
and that circular polarization was detected in 2004 August
with a polarization fraction of $\sim$50\%.

\section{DISCUSSION}

\subsection{The Nature of Variable Radio Emission}

The results of this survey, together with the archival data,
suggest that at least 20\%--30\% of VeLLOs are radio variables.
Out of the 10 target sources,
our survey and the archival data revealed
that three sources showed both detections and nondetections
(for a detection threshold of $\sim$0.05 mJy),
and the signal-to-noise ratios at the time of detections were five or higher.
The 20 day (or shorter) variability timescale
of the IRAM 04191+1522 IRS outburst
is comparable to that of the magnetic flares of YSOs.
For example, the flares of GMR-A showed a timescale of several days
(Bower et al. 2003) to $\sim$15 days (Furuya et al. 2003),
and those of R CrA IRS 5b showed a timescale of $\sim$10 days
(Forbrich et al. 2006; Choi et al. 2008).

The flux variability caused by other kinds of mechanism,
such as the propagation of thermal radio jets,
are much slower, typically taking several years or longer
(e.g., Rodr{\'\i}guez et al. 2000),
and can be ruled out.
If the variable radio emission were thermal radiation from molecular gas,
it should have been easily detectable at shorter wavelengths.
For example, the expected flux densities of the thermal emission
are at least 20 mJy at 3 mm and at least 200 mJy at 1 mm.
The observed values, however, were always much weaker:
undetectable ($<$ 2 mJy) at 3.3 mm (Lee et al. 2005)
and 6--9 mJy at 1.3 mm
(Andr{\'e} et al. 1999; Belloche et al. 2002; Chen et al. 2012).
Therefore, the thermal emission can be ruled out.

The modest amount of circular polarization of L1014 IRS
(Section 3.3; Shirley et al. 2007)
suggests that the detected emission was gyrosynchrotron radiation
from mildly relativistic electrons,
most likely accelerated in magnetic activities.
The nondetection of circular polarization in the IRAM 04191+1522 IRS outburst
does not necessarily contradict the nonthermal nature of the emission
because the amount of polarization depends on many factors
such as geometrical effects
and because the detection upper limit ($\sim$20\%) is rather high.
For example, the polarization fraction of R CrA IRS 5b is $\sim$17\%
(Choi et al. 2009).

Considering the variability timescale and flux level of IRAM 04191+1522 IRS
and the circular polarization of L1014 IRS together,
the variable radio emission of VeLLOs may be coming from magnetic flares.
Therefore, about 20\%--30\% of VeLLOs may be magnetically active,
which may reflect the magnetic properties of the natal clouds
(Nakano et al. 2002).
These observations of VeLLO radio outbursts
may be considered as the detection of light
directly coming from the atmospheres of the youngest stellar objects.
However, given the limited information available,
the nature of the variable radio emission of VeLLOs
is not firmly settled and needs more extensive investigations.
For example, more well-sampled light curves are necessary
to constrain the timescale better.

It should be noted the observations presented in this paper
were designed to detect variabilities on timescales of weeks to months.
It is not clear whether the radio variability of VeLLOs discussed above
is analogous to solar-type flares
that show timescales shorter than an hour
and flux enhancements over orders of magnitude.
The detected variability may be caused
by either several solar-type flares
or a different kind of phenomenon unique to protostars.
To address this issue,
detections of circular polarization and measurements of spectral index
at the time of outbursts
are needed to confirm the nonthermal radiation mechanism,
and nearly continuous observations over several days are necessary.

The undetected VeLLOs were probably magnetically inactive
during the time of observations.
It is also possible
that their protostellar winds may be opaque
and blocking the radio emission from flares,
considering that some VeLLOs can have optically thick winds
(see the range of optical depth in Section 1).
The existence of magnetic activities may apply to more typical protostars
(with luminosities higher than those of VeLLOs),
because magnetic properties do not explicitly depend on the luminosity.
However, their optically thick winds
make it difficult to observationally confirm any magnetic activity.

\subsection{Implications of Magnetic Flares}

In this section we discuss some implications of the radio-variable VeLLOs
assuming that the radio outbursts were caused by magnetic flares.
However, the evidences are not absolutely conclusive,
and the issues discussed below should be revisited
when more and better data are available in the future.

\subsubsection{The Origin of Magnetic Fields}

The detection of VeLLO radio flare suggests
that at least some Class 0 protostars are magnetically active.
The origin of the magnetic fields is an intriguing issue.
The magnetic fields of more evolved objects,
such as T Tauri stars and main-sequence stars, 
may be generated by either solar type or convective dynamos
that require gas circulations such as convection in the stellar interior
(Donati \& Landstreet 2009; Mestel 2012).
Protostars may start to develop a convection zone
when the mass reaches $\sim$0.36 $M_\odot$,
which corresponds to a time soon after the onset of deuterium burning
(Stahler et al. 1980a, 1980b).
However, IRAM 04191+1522 IRS
has a collapse age of $\sim$10,000 yr
and a stellar mass of $\sim$0.05 $M_\odot$ (Andr{\'e} et al. 1999), 
assuming a mass accretion rate of typical protostars.
Considering the low luminosity,
the accretion rate and the stellar mass can be even smaller.
The small age and mass suggest
that this protostar has not developed any convection zone yet.
That is, IRAM 04191+1522 IRS probably has an inert internal region
and may be unable to operate a magnetic dynamo.
The cases of L1014 IRS and L1148 IRS are less certain
because their age and mass are poorly known.
If their masses are smaller than 0.1 $M_\odot$
(Young et al. 2004; Kauffmann et al. 2011),
they also have no convection zone and thus no convective dynamo.

The magnetic fields of pre-main-sequence stars
are often considered to be a combination
of those from the convective dynamo and those from the interstellar space
(Mestel 1965; Tayler 1987).
Since the VeLLOs above cannot operate a convective dynamo,
the source of their magnetic fields may be the fossil fields
that are interstellar magnetic fields of the parent molecular cloud,
dragged into the protostar by the accreting matter.
Therefore, the flares of VeLLOs may be caused by the fossil fields.

However, there is no direct evidence for the existence of fossil fields
in and around protostars so far.
It is in principle possible
that protostars may have an as yet unknown type of internal motion
that can drive a magnetic dynamo.
For example, the fast spin of protostars
may induce internal circulations,
and protostars in a close binary system
may have tidally-induced internal motions.
Theoretical studies of such possibilities
should be helpful but have not been pursued yet.

\subsubsection{High-energy Photons}

While the magnetic flares are most easily detected at radio wavelengths,
most of the energy is released at shorter wavelengths such as X-ray.
Though this issue can be important
in the physics and chemistry of circumstellar material,
there is not much known because VeLLOs have not been detected in X-ray.
Below, we try extrapolating the empirical relations of more evolved objects
to investigate this issue.
However, their applicability to Class 0 protostars
has not been verified observationally,
and the arguments below are speculative.

There is an empirical relation
between radio and X-ray luminosities of large flares:
$L_{\rm XF} / L_{\rm RF} = 10^{15\pm1}$ Hz,
where $L_{\rm XF}$ and $L_{\rm RF}$
are X-ray and radio luminosities during flare events, respectively
(G{\"u}del \& Benz 1993; G{\"u}del 2002).
Many magnetically active objects, including YSOs, follow this relation
(e.g., Bower et al. 2003).
If the 2009 October flare of IRAM 04191+1522 IRS followed this relation,
the X-ray luminosity might have been $\sim$8 $\times$ 10$^{-4}$ $L_\odot$.
By contrast, the X-ray luminosity of a quiescent protostar
can be estimated using another empirical relation:
$L_{\rm XP} / L^*$ = 10$^{-4}$ -- 10$^{-3}$,
where $L_{\rm XP}$ and $L^*$ are X-ray and bolometric luminosities
of an YSO, respectively (G{\"u}del et al. 2007).
Most of the X-ray luminosity of a quiescent YSO
may come from the stellar corona or numerous small flares
(G{\"u}del et al. 2003, 2007).
From the internal luminosity of IRAM 04191+1522 IRS (Dunham et al. 2008),
the X-ray luminosity of the protostar may be $\sim$10$^{-5}$ $L_\odot$.
Comparison of the two X-ray luminosities suggests
that, during large flare events, the magnetic flare
may outshine the protostar by a large factor in the high-energy regime.

Considering that IRAM 04191+1522 IRS
is in active states about half the time,
most of the ionizing photons around this protostar
may be produced by the flares rather than the stellar corona.
The copious amount of high-energy photons from the magnetic flares
may enhance the ionization fraction of the circumstellar material
(Glassgold et al. 2000)
and significantly affect the gas dynamics
such as magnetic breaking and ambipolar diffusion.
Therefore, the magnetic activity can be an important control factor
that may regulate the growth of protostars
(Pudritz \& Silk 1987; Feigelson \& Montmerle 1999; Mestel 2012).

\acknowledgements

We thank Jungyeon Cho for helpful discussions.
M.C. was supported by the Core Research Program
of the National Research Foundation of Korea (NRF)
funded by the Ministry of Science, ICT and Future Planning
of the Korean government (grant No. NRF-2011-0015816).
J.-E.L. was supported by the Basic Science Research Program through NRF
funded by the Ministry of Education of the Korean government
(grant No. NRF-2012R1A1A2044689).
NRAO is a facility of the National Science Foundation
operated under cooperative agreement by Associated Universities, Inc.

\enlargethispage{-10\baselineskip}

\end{document}

%% file: tab1.tex
\begin{deluxetable}{llcccrccc}
\tabletypesize{\small}
\tablecaption{\small Target VeLLOs for the Survey}
\tablewidth{0pt}
\tablehead{
\colhead{Group} & \colhead{Source\tablenotemark{a}}
& \multicolumn{2}{c}{Position\tablenotemark{b}}
& \colhead{$L_{\rm int}$\tablenotemark{c}}
& \colhead{$T_{\rm bol}$\tablenotemark{d}}
& \colhead{Region} & \multicolumn{2}{c}{Distance} \\
\cline{3-4}\cline{8-9}
&& \colhead{$\alpha_{\rm J2000.0}$} & \colhead{$\delta_{\rm J2000.0}$}
& \colhead{($L_\odot$)} & \colhead{(K)} && \colhead{(pc)} & \colhead{Reference}}
\startdata
A & DCE 64 & 03 28 32.57 & $+$31 11 05.3 & 0.03 &  65
           & Perseus     & 250 & Enoch et al. (2006) \\
  & DCE 65 & 03 28 39.10 & $+$31 06 01.8 & 0.02 &  29 \\
  & DCE 81 & 03 30 32.69 & $+$30 26 26.5 & 0.06 &  33 \\
  & DCE  1 & 04 21 56.88 & $+$15 29 46.0 & 0.05 &  27
           & Taurus      & 140 & Kenyon et al. (1994) \\
  & DCE  4 & 04 28 38.90 & $+$26 51 35.6 & 0.03 &  20 \\
B & DCE 24 & 18 16 16.39 & $-$02 32 37.7 & 0.07 &  55
           & CB 130      & 270 & Strai{\v{z}}ys et al. (2003) \\
  & DCE 25 & 18 16 59.47 & $-$18 02 30.5 & 0.07 &  62
           & L328        & 220 & Maheswar et al. (2011) \\
  & DCE 31 & 19 21 34.82 & $+$11 21 23.4 & 0.04 &  24
           & L673        & 240 & Maheswar et al. (2011) \\
  & DCE 32 & 20 40 56.66 & $+$67 23 04.9 & 0.09 & 145
           & L1148       & 300 & Maheswar et al. (2011) \\
  & DCE 38 & 21 24 07.58 & $+$49 59 08.9 & 0.09 &  66
           & L1014       & 260 & Maheswar et al. (2011) \\
\enddata\\
\tablecomments{See Tables 3 and 11 of Dunham et al. (2008)
               for a detailed description of each column.}
\tablenotetext{a}{The source number in Dunham et al. (2008) prefixed with DCE.}
\tablenotetext{b}{Units of right ascension are hours, minutes, and seconds,
                  and units of declination are
                  degrees, arcminutes, and arcseconds.}
\tablenotetext{c}{Internal luminosity ($L_{\rm int}^{70}$)
                  in Table 11 of Dunham et al. (2008).}
\tablenotetext{d}{Bolometric temperature in Table 11 of Dunham et al. (2008).
                  Class 0 protostars have $T_{\rm bol}$ less than 70 K
                  (Chen et al. 1995).}
\end{deluxetable}

%% file: tab2.tex
\begin{deluxetable}{lccrccc}
\tabletypesize{\small}
\tablecaption{\small IRAM 04191+1522 IRS (DCE 1) Parameters}
\tablewidth{0pt}
\tablehead{
\colhead{Date} & \colhead{$F$\tablenotemark{a}}
& \colhead{$\sigma$\tablenotemark{b}}
& \colhead{Beam\tablenotemark{c}} & \colhead{Array\tablenotemark{d}}
& \colhead{$\theta_d$\tablenotemark{e}} & \colhead{$t_i$\tablenotemark{f}} \\
& \colhead{(mJy)} & \colhead{(mJy beam$^{-1}$)} & \colhead{(arcsec)}
&& \colhead{(arcsec)} & \colhead{(minutes)}}
\startdata
1992 Jul 16 & \ldots & 0.016 & 10.9 $\times$ 9.7 & D &    44 & 11 \\
1996 Jul 25 & 0.057  & 0.012 & 12.5 $\times$ 9.3 & D & \phn2 & 58 \\
1997 Nov 12 & 0.112  & 0.017 & 10.5 $\times$ 9.3 & D & \phn2 & 54 \\
1999 Jan 26 & \ldots & 0.015 &  3.2 $\times$ 2.7 & C &    58 & 10 \\
2003 Jan  4 & \ldots & 0.008 &  3.2 $\times$ 2.6 & C &    58 & 31 \\
2009 Oct 18 & 0.142  & 0.009 &  9.6 $\times$ 8.6 & D & \phn0 & 13 \\
2009 Nov  8 & 0.069  & 0.011 &  9.9 $\times$ 8.4 & D & \phn0 & 13 \\
2009 Nov 30 & \ldots & 0.014 & 10.8 $\times$ 9.0 & D & \phn0 & 13 \\
\enddata\\
\tablenotetext{a}{Flux density of the 8.5 GHz continuum,
                  corrected for the primary beam response.}
\tablenotetext{b}{Root-mean-square value of the noise in the map,
                  corrected for the primary beam response.}
\tablenotetext{c}{FWHM of the synthesized beam.}
\tablenotetext{d}{Configuration of the VLA antennas.}
\tablenotetext{e}{Angular distance from the phase-tracking center.
                  The radius of the antenna primary beam is FWHM/2 = 160$''$.}
\tablenotetext{f}{On-source integration time.}
\end{deluxetable}

%% file: tab3.tex
\begin{deluxetable}{lclccrc}
\tabletypesize{\small}
\tablecaption{\small Parameters of Sources Undetected in This Survey}
\tablewidth{0pt}
\tablehead{
\colhead{Source} & \colhead{Alternative} & \colhead{Epoch}
& \colhead{$F$\tablenotemark{a}} & \colhead{$\sigma$\tablenotemark{a}}
& \colhead{Beam} & \colhead{Array} \\
& \colhead{Name} && \colhead{(mJy)} & \colhead{(mJy beam$^{-1}$)}
& \colhead{(arcsec)}}
\startdata
DCE 64 & LAL 38          & 2009 Oct--Nov & \ldots & 0.009
                         &  9.1 $\times$ 8.7 & D \\
DCE 65 & HH 340B         & 2009 Oct--Nov & \ldots & 0.009
                         &  9.2 $\times$ 8.9 & D \\
DCE 81 & \ldots          & 2009 Oct--Nov & \ldots & 0.008
                         &  9.4 $\times$ 9.1 & D \\
DCE  4 & L1521F IRS      & 1999 Mar 6    & \ldots & 0.017
                         & 10.1 $\times$ 9.1 & D \\
       &                 & 2004 Mar 12   & \ldots & 0.009
                         &  2.9 $\times$ 2.7 & C \\
       &                 & 2009 Oct--Nov & \ldots & 0.011
                         &  9.2 $\times$ 8.9 & D \\
DCE 24 & CB 130--1 IRS 1 & 2009 Oct--Nov & \ldots & 0.007
                         & 11.1 $\times$ 8.0 & D \\
DCE 25 & L328 IRS        & 2009 Oct--Nov & \ldots & 0.010
                         & 13.6 $\times$ 7.3 & D \\
DCE 31 & L673--7 IRS     & 2009 Oct--Nov & \ldots & 0.012
                         & 10.3 $\times$ 8.4 & D \\
DCE 32 & L1148 IRS       & 2005 Jul 20   & 0.060  & 0.008
                         &  3.3 $\times$ 2.9 & C \\
       &                 & 2009 Oct--Nov & \ldots & 0.012
                         & 11.0 $\times$ 8.0 & D \\
DCE 38\tablenotemark{b}
       & L1014 IRS       & 2004 Jul 1    & 0.103  & 0.015
                         &  5.9 $\times$ 5.9 & D \\
       &                 & 2005 Dec 5    & 0.119  & 0.020
                         &  9.5 $\times$ 5.9 & D \\
       &                 & 2009 Oct--Nov & \ldots & 0.009
                         &  9.6 $\times$ 8.4 & D \\
\enddata\\
\tablenotetext{a}{The VeLLO targets were
                  close to the corresponding phase-tracking centers,
                  and no primary-beam correction was necessary.}
\tablenotetext{b}{The parameters of the 2004/2005 observations of L1014 IRS
                  are from Shirley et al. (2007).}
\end{deluxetable}

%% file: fig1.tex
\begin{figure*}[!t]
\epsscale{2.0}
\plotone{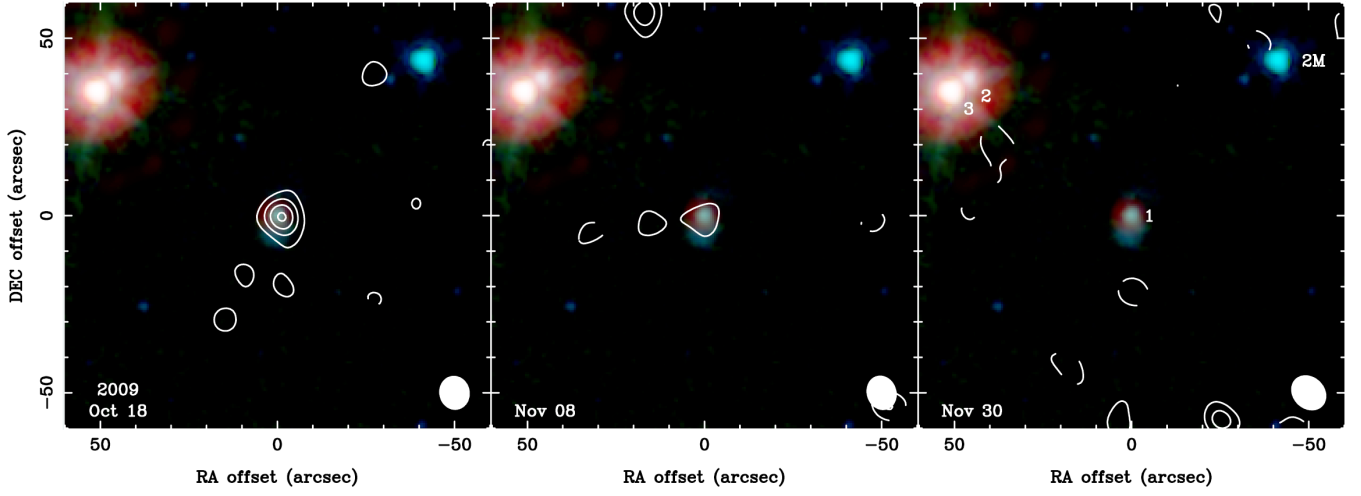}
\caption{
Contour maps of the 8.5 GHz continuum emission
toward the IRAM 04191+1522 region in 2009.
The date of observation is labeled in each panel.
The contour levels are 1, 2, 3, and 4 $\times$ 0.03 mJy beam$^{-1}$.
Dashed contours are for negative levels.
These maps are not corrected for the primary beam response.
Shown in the bottom right-hand corner of each panel is the synthesized beam.
The background color-composite image
shows the infrared maps
of the 3.6 (blue), 4.5 (cyan), 5.8 (green), and 24 (red) $\mu$m bands
from {\it Spitzer Space Telescope} (Dunham et al. 2006).
The infrared sources are IRAS 04191+1523 A (DCE 3), IRAS 04191+1523 B (DCE 2),
IRAM 04191+1522 IRS (DCE 1),
and Two Micron All Sky Survey (2MASS) J04215402+1530299,
from left to right (Dunham et al. 2006, 2008).
Labeled in the last panel are the DCE numbers and `2M' for the 2MASS source.}
\end{figure*}

%% file: fig2.tex
\begin{figure}[!t]
\epsscale{1.0}
\plotone{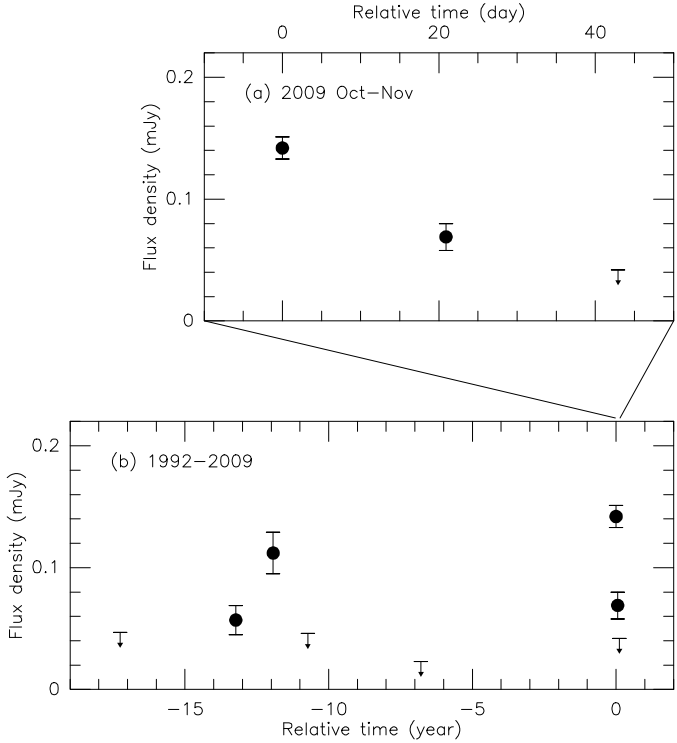}
\caption{
Temporal variability of the 8.5 GHz continuum of IRAM 04191+1522 IRS.
(a) Light curve in 2009 October--November.
(b) Light curve in 1992--2009.
For the epochs of nondetection, 3$\sigma$ upper limits are marked.
The horizontal axis shows the time of observations
relative to the first observing run in 2009.}
\end{figure}

%% file: fig3.tex
\begin{figure}[!t]
\epsscale{1.0}
\plotone{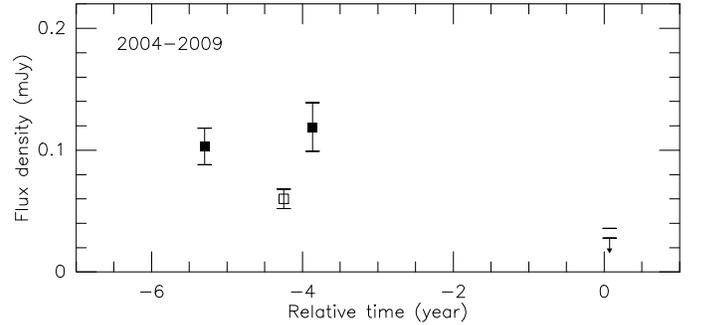}
\caption{
Temporal variability of the 8.5 GHz continuum
of L1148 IRS (open square) and L1014 IRS (solid squares).
For the observations in 2009, 3$\sigma$ upper limits are marked.
The horizontal axis shows the time of observations
relative to Julian date 2455122.98 (the same as in Figure 2).}
\end{figure}